\documentclass[pra,aps,superscriptaddress,showpacs,twocolumn]{revtex4}

\usepackage{bm}
\usepackage{amssymb}
\usepackage{graphicx}

\begin{document}

\title{Experimental demonstration of optimal universal asymmetric quantum cloning 
of~polarization states of single photons by partial symmetrization}

\author{Anton\'{\i}n \v{C}ernoch}
\affiliation{Joint Laboratory of Optics of Palack\'{y} University
and Institute of Physics of Academy of Sciences of the Czech
Republic, 17. listopadu 50A, 779\,07 Olomouc, Czech Republic}

\author{Jan Soubusta}
\affiliation{Joint Laboratory of Optics of Palack\'{y} University
and Institute of Physics of Academy of Sciences of the Czech
Republic, 17. listopadu 50A, 779\,07 Olomouc, Czech Republic}

\author{Lucie \v{C}elechovsk\'{a}}
\affiliation{Department of Optics, Palack\'y
University, 17.~listopadu 12, 779\,00 Olomouc, Czech~Republic}

\author{Miloslav Du\v{s}ek}
\affiliation{Department of Optics, Palack\'y
University, 17.~listopadu 12, 779\,00 Olomouc, Czech~Republic}

\author{Jarom{\'\i}r Fiur\'{a}\v{s}ek}
\affiliation{Department of Optics, Palack\'y
University, 17.~listopadu 12, 779\,00 Olomouc, Czech~Republic}

\begin{abstract}

We report on experimental implementation of the optimal universal asymmetric $1\rightarrow 2$ quantum cloning machine for qubits encoded into polarization states of single photons. Our linear optical  machine performs asymmetric cloning by partially symmetrizing the input polarization state of signal photon and a blank copy idler photon prepared in a maximally mixed state. We show that the employed method of measurement of mean clone fidelities exhibits strong resilience to imperfect calibration of the relative efficiencies of single-photon detectors used in the experiment. Reliable characterization of the quantum cloner is thus possible even when precise detector calibration is difficult to achieve.

\end{abstract}

\maketitle

\section{Introduction}

Unlike classical information, quantum information cannot be freely copied. 
The famous quantum no-cloning theorem \cite{Wootters82,Dieks82} states that it is impossible to perfectly copy non-orthogonal quantum
states. This puts a nontrivial bound on the ability to share quantum information among
several parties which can be explored in quantum information processing applications such as
quantum cryptography \cite{Gisin02,Scarani09}. Although perfect quantum copying is forbidden, one can nevertheless
attempt to copy the quantum states in the optimal approximate manner and this 
has been a subject of numerous theoretical and experimental
studies  during the recent years \cite{Scarani05}. 
Optimal quantum cloners have been identified for various sets of
input states and some of them have been demonstrated experimentally \cite{Cummins02,Linares02,Ricci04,Irvine04,Du05,Cernoch06,Chen07,Bartuskova07,Soubusta08}. 

Of particular interest is the optimal universal quantum cloner that should copy equally well all states from the underlying Hilbert space \cite{Buzek96,Gisin97,Linares02}. The universal quantum cloning can be accomplished by a symmetrization of a state to be cloned and a maximally mixed state of blank copies \cite{Werner98}. This insight has led to the experimental realization of the quantum cloning of polarization states of photons via bunching on a balanced beam splitter \cite{Ricci04,Irvine04,Khan04}. This procedure
provides two clones with identical reduced density matrices and fidelities. However, one can
more generally consider  asymmetric quantum cloning machine that produces two clones
with different fidelities $F_A$ and $F_B$ such that the trade-off between $F_A$ and $F_B$ is
optimal \cite{Niu98,Cerf00,Iblisdir05}. Optimal asymmetric phase-covariant cloning has been demonstrated 
in a nuclear magnetic resonance experiment \cite{Chen07} and in two optical experiments involving either optical fibers and qubits encoded into path of single photons \cite{Bartuskova07} or bulk optics and polarization encoding \cite{Soubusta08}. By contrast, so far only a single experimental realization of the universal asymmetric cloner has been reported. In that experiment, asymmetric cloning of polarization states of single photons was accomplished via partial quantum teleportation \cite{Zhao05,Filip04} which required detection of four-photon coincidences resulting in low coincidence rate. The performance of the cloner was thus characterized only for a rather limited set of three different input states. 

Here we report on the experimental demonstration of optimal universal asymmetric cloning via
the partial symmetrization of a two-photon polarization state. Our linear-optical 
partial symmetrizer consists in  a Mach-Zehnder interferometer with additional beam splitters in its arms \cite{Fiurasek08,Cernoch09}.
The asymmetry of such cloner can be easily tuned by changing the transmittance of one arm of
the interferometer using variable attenuator. 
We characterize the performance of the cloner for various degrees of asymmetry by the mean fidelities of the two clones which
are experimentally determined as averages of clone fidelities for input states forming three mutually unbiased bases. 

An important issue in the experiments where multiphoton coincidences are measured is the
proper calibration of the relative detection efficiencies of the single photon detectors. We show that, interestingly, our
method of measurement of the mean clone fidelities is inherently robust with respect to the
errors in detection efficiency calibration. Even errors of the order of several ten per-cent
can be tolerated as the resulting systematic error in determination of the mean clone
fidelities is of the order of $0.5\%$ or even smaller. This result implies that
reliable estimates of certain parameters such as average fidelities can be obtained even when
it is difficult to perfectly calibrate the detectors.

The rest of the paper is structured as follows. In Section II we describe the experimental
setup and the measurement procedure. The experimental results are
presented and discussed in Sec. III, where it is shown that our method of determination of the
mean fidelities is inherently robust with respect to imperfect detector efficiency calibration.
Finally, Section IV contains a brief summary and conclusions.

\section{Experimental setup}

The experimental setup for partial symmetrization of the polarization state of two photons 
is depicted in Fig. 1 and is  described in detail in Refs. \cite{Cernoch08,Cernoch09}. In brief, correlated photon pairs
are produced by the process of Type-I degenerate spontaneous parametric downconversion (SPDC) in a
nonlinear crystal LiIO$_3$ pumped by a cw Kr-ion laser with wavelength 413 nm. The signal and idler
photons are spatially filtered by coupling them into single-mode fibers. The input polarization
states of the photons are prepared by means of half- and quarter-wave plates. The photons then
interfere in a Mach-Zehnder interferometer supplemented by two additional beam-splitters and one 
neutral density filter (NDF) which serves as a polarization-insensitive attenuator. The output
states of the photons are measured by means of two polarization-detection blocks consisting of a
sequence of quarter- and half-waveplates, polarizing beam splitter and two single photon
detectors. The device operates in the coincidence basis and the partial symmetrization is successful
if a single photon is detected by each block.

\begin{figure}[!t!]
\centerline{\includegraphics[width=0.95\linewidth]{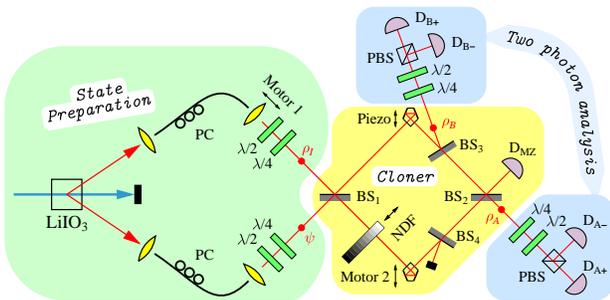}}
\caption{(Color online) Experimental setup: PC - fiber polarization controller,
BS - non-polarizing plate beam splitter, NDF - neutral density
filter, PBS - polarizing cube beam splitter, $\lambda/2$ and
$\lambda/4$ - wave plates, D - describes a set composed of cut-off
filter, collimating lens, single-mode fiber and avalanche
photodiode, D$_{\mathrm{MZ}}$ - auxiliary single photon detector used for stabilization of the interferometer.\label{setup}}
\end{figure}

As shown in Ref. \cite{Fiurasek08,Cernoch09}, the modified Mach-Zehnder interferometer conditionally implements a partial symmetrization operation on the polarization state of the two photons,
\begin{equation}
V_S=\Pi_{+}+t\,\Pi_{-}.
\label{VS}
\end{equation}
Here $\Pi_{-}=|\Psi^{-}\rangle\langle \Psi^{-}|$ is the projector onto the one-dimensional
anti-symmetric subspace of the Hilbert space of two qubits $\mathcal{H}$, $|\Psi^{-}\rangle=\frac{1}{\sqrt{2}}(|HV\rangle-|VH\rangle)$ is the
singlet Bell state and $|H\rangle$ and $|V\rangle$ denote horizontal and vertical polarization state of a single photon, respectively. Similarly, $\Pi_{+}=\openone-\Pi_{-}$ denotes projector onto the symmetric subspace of $\mathcal{H}$, $\openone$ is the identity operator, and $t$ denotes the amplitude transmittance of NDF.  In the experiment, the interferometer is actively stabilized at zero phase shift between the two arms which guarantees that $t$ in Eq. (\ref{VS}) is real and positive. The partial 
symmetrization begins by a two-photon interference  on the balanced beam splitter BS$_1$.
Photons in anti-symmetric singlet Bell state $|\Psi^{-}\rangle$ anti-bunch and each photon
propagates through one arm of the MZ interferometer. On the other hand, if the photons are in a
symmetric polarization state, then they bunch and both must propagate through the upper  arm of
the interferometer in order to reach the output ports. An attenuator placed in the lower
interferometer arm thus selectively attenuates the anti-symmetric part of the two-photon
polarization state.

In the implementation of the optimal universal asymmetric $1\rightarrow 2$ cloning operation, the signal
photon carries the state $|\psi\rangle$ that should be copied, while the idler photon represents
a blank copy and is prepared in a maximally mixed state. In the experiment, this is achieved by 
randomly preparing the idler photon in one of the six polarization states $|H\rangle$,
$|V\rangle$, $|D\rangle=\frac{1}{\sqrt{2}}(|H\rangle+|V\rangle)$, $|A\rangle=\frac{1}{\sqrt{2}}(|H\rangle-|V\rangle)$, $|R\rangle=\frac{1}{\sqrt{2}}(|H\rangle+i|V\rangle)$ and $|L\rangle=\frac{1}{\sqrt{2}}(|H\rangle-i|V\rangle)$, with equal probability $\frac{1}{6}$.
The input two-photon state can thus be written as  $\rho_I \otimes \psi$, where $\psi=|\psi\rangle\langle\psi|$
denotes the density matrix of a pure state $|\psi\rangle$ and $\rho_I=\frac{1}{2}(|H\rangle\langle H|+|V\rangle\langle V|)$ stands for the density matrix of a maximally mixed state.
 The output state of the partial symmetrizer reads
\begin{equation}
\rho_{\mathrm{out}}= V_S \left(\rho_I\otimes  \psi\right) V_S^\dagger= (\Pi_{+}+t\Pi_{-})
\rho_I \otimes \psi(\Pi_{+}+t\Pi_{-}).
\label{rhoout}
\end{equation}
After some algebra we find the following expressions for the normalized reduced density matrices of the two clones A
and B, $\rho_A=\mathrm{Tr}_{B}[\rho_{\mathrm{out}}]/\mathrm{Tr[\rho_{\mathrm{out}}]}$, 
$\rho_B=\mathrm{Tr}_{A}[\rho_{\mathrm{out}}]/\mathrm{Tr[\rho_{\mathrm{out}}]}$,
\begin{eqnarray}
\rho_A=\frac{1}{2(3+t^2)}[(5-2t+t^2)\psi+(1+t)^2\psi_\perp], \nonumber \\
\rho_B=\frac{1}{2(3+t^2)}[(5+2t+t^2)\psi+(1-t)^2\psi_\perp].  
\label{rhoAB}
\end{eqnarray}
The fidelities of the two clones, $F_A=\langle \psi|\rho_A|\psi\rangle$, 
$F_B=\langle \psi|\rho_B|\psi\rangle$, can be immediately determined from Eq. (\ref{rhoAB}) and we obtain,
\begin{equation}
F_A=\frac{5-2t+t^2}{2(3+t^2)}, \qquad F_B=\frac{5+2t+t^2}{2(3+t^2)}.
\label{FAB}
\end{equation}
We can see that the transmittance $t$ of the NDF  controls the asymmetry of the cloner, i.e. the
fidelities of the two clones. It can be verified that the fidelities (\ref{FAB}) satisfy the optimal
trade-off between $F_A$ and $F_B$ achievable by the universal $1\rightarrow 2$ cloning machine \cite{Niu98,Cerf00},
\begin{equation}
(1-F_A)(1-F_B)=\left(F_A+F_B-\frac{3}{2}\right)^2.
\label{FABtradeoff}
\end{equation}
This confirms that the scheme in Fig.~1 implements the optimal asymmetric universal cloning
operation. 

Note that the scheme can be straightforwardly extended 
such as to produce, in addition to the two optimal clones, also the anti-clone \cite{Ricci04,Irvine04}. 
Instead of a single idler photon in a maximally mixed polarization state one would have to employ
 a pair of photons prepared in a maximally entangled polarization singlet
Bell state. One photon from this maximally entangled
pair whould replace the idler photon and be 
injected into the partial symmetrizer together with a signal photon whose polarization state
should be cloned. The two partially symmetrized photons would carry the clones while the
remaining photon from the maximally entangled pair would represent the anti-clone. However, 
such an extended setup would involve
three-photon coincidence detection and would require pumping of the nonlinear crystal with
femtosecond pulsed laser in order to achieve the necessary temporal synchronization and overlap
of photons' wavepackets \cite{Ricci04,Irvine04}.

In the experiment, we measure the clone fidelity for six different input states forming three mutually
unbiased bases $\{|H\rangle,|V\rangle\}$, $\{|D\rangle,|A\rangle\}$, and
$\{|R\rangle,|L\rangle\}$. For any basis $|\psi\rangle,|\psi_\perp\rangle$ we first set the
waveplates in the detection blocks such that clicks of detectors $D_{A+}$ and $D_{B+}$ represent
projection of photon onto polarization state $|\psi\rangle$ while clicks of $D_{A-}$ and $D_{B-}$
indicate projection onto state $|\psi_\perp\rangle$. We measure  four two-photon
coincidence rates $C_{jk}$ of simultaneous clicks of $D_{Aj}$ and $D_{Bk}$, where $j,k\in\{+,-\}$.
For instance, $C_{+-}$ always denotes number 
of simultaneous clicks of detectors $D_{A+}$ and $D_{B-}$. 
For input signal state $|\psi\rangle$ the clicks of detector $D_{A+}$ ($D_{B+}$) herald 
successful preparation of clone A (B) in state $|\psi\rangle$ and the 
fidelities of the two clones can thus be calculated as \cite{Cernoch06}
\begin{eqnarray}
f_{A,\mathrm{exp}}(\psi)=\frac{C_{++}+C_{+-}}{C_{++}+C_{+-}+C_{-+}+C_{--}}, \nonumber \\
f_{B,\mathrm{exp}}(\psi)=\frac{C_{++}+C_{-+}}{C_{++}+C_{+-}+C_{-+}+C_{--}}.
\label{fABexppsi}
\end{eqnarray}
We then change the input state of the signal photon to $|\psi_\perp\rangle$, while leaving the
detection blocks unchanged. We again measure the four coincidence rates $C_{jk}$. This time a
successful preparation of clone A (B) in state $|\psi_\perp\rangle$ is heralded by clicks of
detector  $D_{A-}$ ($D_{B-}$) and the fidelities of the two clones are given by
\begin{eqnarray}
f_{A,\mathrm{exp}}(\psi_{\perp})=\frac{C_{--}+C_{-+}}{C_{++}+C_{+-}+C_{-+}+C_{--}}, \nonumber
\\
f_{B,\mathrm{exp}}(\psi_{\perp})=\frac{C_{--}+C_{+-}}{C_{++}+C_{+-}+C_{-+}+C_{--}}.
\label{fABexppsiperp}
\end{eqnarray}
Note that throughout the paper we shall use lowercase $f$ to denote fidelities of clones of single states  and uppercase $F$ to denote mean fidelities of the cloner.
The mean fidelity of each clone is calculated by averaging the clone fidelities obtained for the 
three mutually unbiased bases, 
\[
\bar{F}_A=\frac{1}{6}\sum_{j}f_{A,\mathrm{exp}}(j), \qquad \bar{F}_B=\frac{1}{6}\sum_{j}f_{B,\mathrm{exp}}(j),
\]
where the summation runs over $j=|H\rangle,$ $|V\rangle,$ $|D\rangle,$ $|A\rangle,$ $|R\rangle,$ $|L\rangle$.

\begin{figure}[!t!]
\centerline{\includegraphics[width=0.98\linewidth]{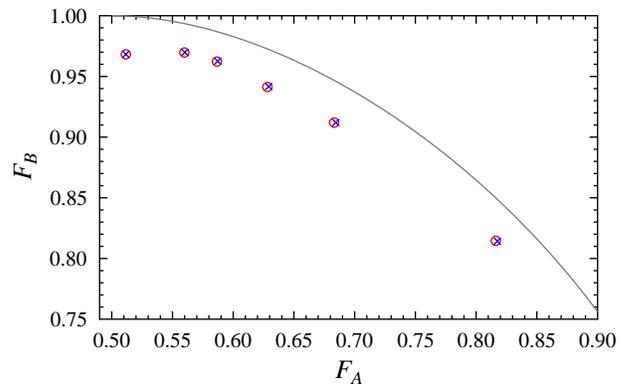}}
\caption{(Color online) Trade-off between the fidelities of the two clones $F_A$ and $F_B$. The symbols represent  pairs of fidelities determined from the experimental data  before calibration (red circles) and after calibration (blue crosses) of the relative detection efficiency, see main text.  The solid line indicates the best possible trade-off achievable by the optimal universal asymmetric quantum cloner.}
\end{figure} 

\section{Results and discussion}

The performance of the cloner was experimentally tested for six different degrees of asymmetry
$t=\sqrt{\frac{n}{5}}$, $n=0,1,2,3,4,5$. The resulting mean fidelities of the two clones are plotted in Fig.
2 together with the optimal theoretical trade-off curve specified by formula (\ref{FAB}). We can
see that the cloner exhibits good performance and  the experimental data follow the theoretical
curve. The measured fidelities lie somewhat below the optimal theoretical bound
which can be attributed mainly to the imperfect visibility of the single- and two-photon interference 
in our setup.

\begin{figure}[!t!]
\centerline{\includegraphics[width=0.85\linewidth]{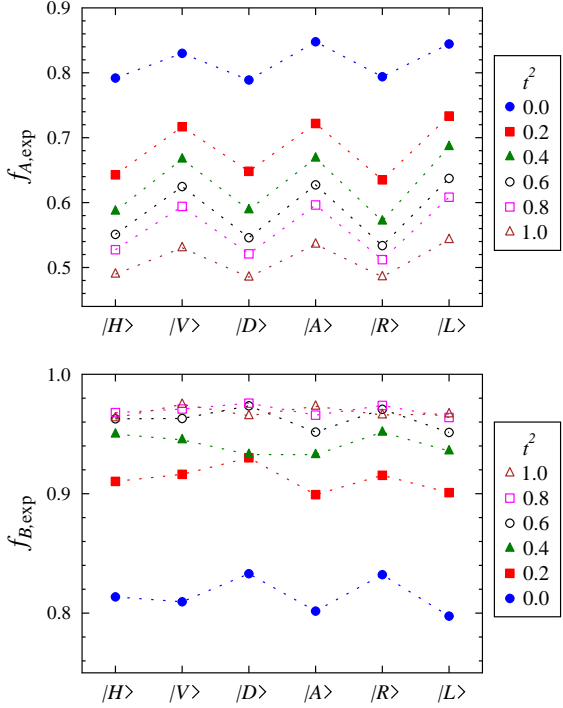}}
\caption{(Color online) Experimentally determined fidelities of the first clone $f_{A,\mathrm{exp}}$ (upper panel) and second clone $f_{B,\mathrm{exp}}$ (lower panel) are plotted for each of the six input states $|\psi\rangle$ and for the six different degrees of asymmetry (see text) which are indicated by different colors and symbols.}
\end{figure}

Figure 3 shows the fidelities of the first and second clones for each of the six input states.
Since the cloner should be universal, one expects that for a given degree of asymmetry the
fidelity should be the same for all inputs. However, a saw-like oscillatory dependence of
$f_{A,\mathrm{exp}}$ and $f_{B,\mathrm{exp}}$ on the input state is clearly visible in Fig. 3. Namely, for each basis
$|\psi\rangle,|\psi_{\perp}\rangle$ the fidelity of the clone A of state $|\psi\rangle$ is lower than
the fidelity of the clone A of $|\psi_\perp\rangle$, while an opposite behavior is found for
clone B. Taking into account our method of fidelity measurement, this can be explained by an 
imperfect calibration of the detection efficiencies of the employed APDs. In the experiment, we performed auxiliary measurements from which the detection efficiencies were estimated. The experimental data were then properly re-scaled. However, as Fig. 3 reveals this procedure did not completely compensate for the different
efficiencies.

Since we measure coincidence rates between one detector $D_{Aj}$ and one detector
$D_{Bk}$, what matters are only the relative efficiencies of the two detectors in each detection
block. We denote by $\eta_A$ the ratio of detection efficiencies of $D_{A-}$ and $D_{A+}$, and
$\eta_B$ is defined similarly. The non-unit relative detection efficiencies can be accounted for by properly rescaling the coincidence rates,
\begin{eqnarray}
C_{++} & \rightarrow & \eta_A\eta_B C_{++}, \nonumber \\
C_{+-} & \rightarrow & \eta_A C_{+-}, \nonumber \\
C_{-+} & \rightarrow & \eta_B C_{-+}, \nonumber \\
C_{--} & \rightarrow &  C_{--}.
\label{Crescaled}
\end{eqnarray}
We can attempt to calibrate the detectors by determining $\eta_A$ and $\eta_B$ that minimize the spread of the fidelity for a fixed asymmetry of the cloner. We have chosen to minimize the fidelity variance defined as
\begin{equation}
\Delta \bar{F}_A^2
=\frac{1}{6}\sum_{j}f_{A,\mathrm{exp}}^2(j)-\frac{1}{36}\left[\sum_{j}f_{A,\mathrm{exp}}(j)\right]^2,
\label{Fvariance}
\end{equation}
where the summation is again performed over the six input states, $j=|H\rangle$, $|V\rangle$, $|D\rangle$, $|A\rangle$, $|R\rangle$, $|L\rangle$. Numerical
minimization of the fidelity variance yields the following values of the relative detection
efficiencies,
\begin{equation}
\eta_{A}=1.046, \qquad \eta_{B}=0.840.
\label{etacalibration}
\end{equation} 
The resulting experimental fidelities obtained from re-calibrated data (\ref{Crescaled}) are plotted in Fig. 4. 
We can see that for all degrees of asymmetry the calibration significantly reduced the spread of the fidelities, which are now practically constant and independent of the state to be cloned. This strongly indicates that this calibration
yields reliable values of $\eta_A$ and $\eta_B$.

\begin{figure}[!t!]
\centerline{\includegraphics[width=0.85\linewidth]{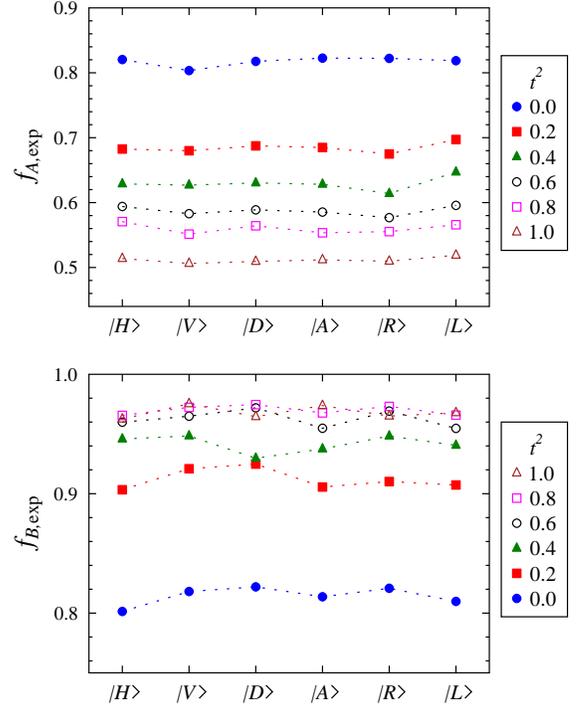}}
\caption{(Color online) The same as Fig. 3 but the plot shows fidelities determined from the experimental data after the calibration using relative detector efficiencies given by Eq. (\ref{etacalibration}). }
\end{figure}

Remarkably, the mean fidelities are almost unaffected by the calibration, as can be seen from
Fig. 2. 
This resilience is a direct consequence of our method of measurement of the clone fidelities. 
The imperfect calibration results in underestimation of the fidelity $f_A(\psi)$ and overestimation 
of $f_A(\psi_{\perp})$ (and vice versa for clone B). The mean fidelity is proportional to $f(\psi)+f(\psi_{\perp})$ and this
averaging significantly reduces the influence of the imperfect calibration on the final result. 
To show this quantitatively, we explicitly write down the dependence of the clone fidelities on
$\eta_{A}$ and $\eta_{B}$.  Consider ideal optimal asymmetric universal quantum cloner 
with input state $|\psi\rangle$. The diagonal elements of the density matrix of the two
clones in basis $|\psi\rangle,|\psi_\perp\rangle$ can be expressed as follows,
\begin{eqnarray}
\rho_{\psi\psi,\psi\psi}&=&P, \nonumber \\
\rho_{\psi\psi_\perp,\psi\psi_\perp}&=&F_A-P, \nonumber \\
\rho_{\psi_\perp\psi,\psi_\perp\psi}&=&F_B-P, \nonumber \\
\rho_{\psi_\perp\psi_\perp,\psi_\perp\psi_\perp}&=&1+P-F_A-F_B.
\label{rhoPF}
\end{eqnarray}
where the parameter $P=2/(3+t^2)$ was introduced to simplify the notation. Note, that the 
formula (\ref{rhoPF}) does not describe only optimal universal quantum cloner but also more general class of covariant quantum machines characterized by three independent parameters $F_A$, $F_B$ and $P$.
Our analysis given below thus applies to all such machines.

The coincidence rates measured by detectors exhibiting relative detection efficiencies $\eta_A$ and $\eta_B$  are given by
\begin{eqnarray}
C_{++}&=&N\rho_{\psi\psi,\psi\psi}, \nonumber \\
C_{+-}&=&N\eta_B \rho_{\psi\psi_\perp,\psi\psi_\perp}, \nonumber \\
C_{-+}&=&N\eta_A\rho_{\psi_\perp\psi,\psi_\perp\psi}, \nonumber \\
C_{--}&=&N\eta_A\eta_B\rho_{\psi_\perp\psi_\perp,\psi_\perp\psi_\perp},
\label{Crho}
\end{eqnarray}
where $N$ is a constant specifying the overall coincidence rate. 
On inserting the expressions (\ref{rhoPF}) and (\ref{Crho}) into
Eq. (\ref{fABexppsi}) we obtain formula for the fidelity of clone A of state $|\psi\rangle$,
\begin{widetext}
\begin{equation}
f_A(\psi)=\frac{P+(F_A-P)\eta_B}{P+(F_A-P)\eta_B+(F_B-P)\eta_A+(1+P-F_A-F_B)\eta_A\eta_B}.
\label{fApsi}
\end{equation}
When calculating the measured fidelity for the orthogonal input state $|\psi_\perp\rangle$ 
we must take into account that due to our measurement method the role of the detectors is reversed with respect to input state
$|\psi\rangle$. Formally, this means that one can still use Eq. (\ref{fApsi}) but has to replace $\eta_A
\rightarrow \eta_A^{-1}$ and $\eta_B\rightarrow \eta_B^{-1}$. We thus have
\begin{equation}
f_A(\psi_\perp)=\frac{P\eta_A\eta_B+(F_A-P)\eta_A}{P\eta_A\eta_B+(F_A-P)\eta_A+(F_B-P)\eta_B+1+P-F_A-F_B}.
\label{fApsiperp}
\end{equation}
\end{widetext}
The mean fidelity of the clone can then be calculated as average of fidelities (\ref{fApsi}) 
and (\ref{fApsiperp}),
\begin{equation}
\bar{F}_A=\frac{1}{2}[f_A(\psi)+f_A(\psi_\perp)].
\label{fAbar}
\end{equation}
In order to gain insight into the dependence of $\bar{F}_A$ on the relative efficiencies $\eta_A$
and $\eta_B$ we introduce new parameters $\epsilon_A$ and $\epsilon_B$ that directly quantify the
relative efficiency mismatch, 
\begin{equation}
\eta_A=1+\epsilon_A, \qquad \eta_B=1+\epsilon_B,
\label{etaepsilon}
\end{equation}
and expand $\bar{F}_A$ in Taylor series in $\epsilon_A$ and $\epsilon_B$ up to the second order. 
After some algebra we obtain
\begin{eqnarray}
\bar{F}_A& \approx & F_A 
+\frac{1}{2}F_A(1-F_A)(1-2F_A)\epsilon_A^2 \nonumber \\
& & +(2F_A-1)(F_AF_B-P)\epsilon_A \epsilon_B \nonumber \\
& & +\frac{1}{2}(P-F_AF_B)(1-2F_B)\epsilon_B^2.
\label{fAbarexpansion}
\end{eqnarray}
The terms linear in $\epsilon$ stemming from $f_A(\psi)$ and $f_A(\psi_\perp)$ exactly cancel each
other so the estimation error depends only quadratically on $\epsilon_A$ and
$\epsilon_B$. Moreover, 
the coefficients specifying the quadratic error term in Eq. (\ref{fAbarexpansion}) are typically very small. 
To give a concrete example let us consider optimal symmetric universal quantum cloner. We then have
\begin{equation}
P=\frac{2}{3}, \qquad F_A=F_B=\frac{5}{6}. 
\end{equation}
When we insert these values into Eq. (\ref{fAbarexpansion}) we obtain 
\begin{equation}
\Delta \bar{F}_A \equiv \bar{F}_A-F_A \approx -\frac{5}{108}\epsilon_A^2+\frac{1}{54}\epsilon_A\epsilon_B
+\frac{1}{108}\epsilon_B^2.
\label{deltafAbar}
\end{equation}
We can bound the absolute value of the error by a maximum eigenvalue of a $2\times 2$ matrix that
describes the quadratic form in Eq. (\ref{deltafAbar}), 
\begin{equation}
|\Delta \bar{F}_A| \lesssim \frac{2+\sqrt{10}}{108}(|\epsilon_A|^2+|\epsilon_B|^2).
\end{equation}
This implies that for instance a $10\%$ error in relative calibration, $|\epsilon_A|=0.1$, results in only a
very small error in estimation of the cloning fidelity, $\Delta \bar{F}_A \approx 0.05\%$. A
calculation based on the exact formulas (\ref{fApsi}) and (\ref{fApsiperp}) confirms this result.
The analysis was presented here for clone $A$ but similar formulas and conclusions hold also for the clone B, one only has to properly interchange labels $A$ and $B$ in the relevant equations.

\section{Summary}

In summary, we have demonstrated optimal universal asymmetric cloning of polarization states of single photons by partial symmetrization. The linear optical symmetrizer combines single- and two-photon interference in a modified Mach-Zehnder interferometer and the asymmetry of the cloner can be easily modified by inserting a variable attenuator in one arm of the interferometer. We have shown that the employed method of measurement of the mean clone fidelities has a built-in resilience against imperfect calibration of the efficiencies of the single photon detectors. This measurement procedure thus allowed a highly reliable characterization of the cloner and can be  potentially useful also for characterization of other devices such as linear optical quantum gates.

\acknowledgments

This research was supported by the projects LC06007,
1M06002 and MSM6198959213 of the Ministry of Education
of the Czech Republic and by GACR (GA202/09/0747).

\end{document}